\documentclass[a4paper,fleqn,usenatbib]{mnras}
\usepackage[T1]{fontenc}
\usepackage{ae,aecompl}
\usepackage{graphicx}	
\usepackage{amsmath}	
\usepackage{amssymb}
\def\lya{Ly$\alpha$~}
\usepackage{color}

\title[21~cm models]{Large 21~cm signals from AGN-dominated reionization}

\author[Kulkarni et al.]{{Girish Kulkarni$^{1}$\thanks{Email:
      kulkarni@ast.cam.ac.uk}, Tirthankar Roy Choudhury$^{2}$, Ewald
    Puchwein$^1$} \newauthor{and Martin G. Haehnelt$^1$} \\ $^1$Institute
  of Astronomy and Kavli Institute of Cosmology, University of
  Cambridge, Madingley Road, Cambridge CB3 0HA, UK \\ $^2$National
  Centre for Radio Astrophysics, Tata Institute of Fundamental
  Research, Post Bag 3, Ganeshkhind, Pune 411007, India}

\date{Accepted ---. Received ---; in original form ---}

\pubyear{2016}

\begin{document}
\label{firstpage}
\pagerange{\pageref{firstpage}--\pageref{lastpage}}
\maketitle

\begin{abstract}
  We present predictions for the spatial distribution of 21~cm
  brightness temperature fluctuations from high-dynamic-range
  simulations for AGN-dominated reionization histories that have been
  tested against available \lya and CMB data.  We model AGN by
  extrapolating the observed M$_\mathrm{bh}$--$\sigma$ relation to
  high redshifts and assign them ionizing emissivities consistent with
  recent UV luminosity function measurements. We assess the
  observability of the predicted spatial 21~cm fluctuations in the
  late stages of reionization in the limit in which the hydrogen 21~cm
  spin temperature is significantly larger than the CMB temperature.
  Our AGN-dominated reionization histories increase the variance of
  the 21~cm emission by a factor of up to ten compared to similar
  reionization histories dominated by faint galaxies, to values close
  to 100 mK$^2$ at scales accessible to experiments ($k \lesssim 1$
  cMpc$^{-1}h$). This is lower than the sensitivity reached by ongoing
  experiments by only a factor of about two or less.  When
  reionization is dominated by AGN, the 21~cm power spectrum is
  enhanced on all scales due to the enhanced bias of the clustering of
  the more massive haloes and the peak in the large scale 21~cm power
  is strongly enhanced and moved to larger scales due to bigger
  characteristic bubble sizes.  AGN dominated reionization should be
  easily detectable by LOFAR (and later HERA and SKA1) at their design
  sensitivity, assuming successful foreground subtraction and
  instrument calibration.  Conversely, these could become the first
  non-trivial reionization scenarios to be ruled out by 21~cm
  experiments, thereby constraining the contribution of AGN to
  reionization.
\end{abstract}


\section{Introduction}
\label{sec:intro}

Hydrogen reionization is generally thought to occur at redshifts
$z\sim 6$--$15$ by Lyman continuum photons that are widely believed to
be produced by young stars in low-mass galaxies
\citep{2015MNRAS.454L..76M}.  However, the idea that active galactic
nuclei (AGN) could have been the dominant source of ionizing radiation
during the epoch of reionization has recently gained traction again
\citep{2015A&A...578A..83G, 2015ApJ...813L...8M, 2015MNRAS.453.2943C,
  2016MNRAS.457.4051K, 2017MNRAS.465.3429C, 2016arXiv160602719M,
  2016arXiv160706467D}.

The resurgence of AGN as credible source of ionising photons at high
redshift is due to a number of recent developments.  First, the
claimed discovery of 19 low-luminosity ($M_{1450}>-22.6$) AGN between
redshifts $z=4.1$ and $6.3$ by \citet{2015A&A...578A..83G} using a
novel X-ray/NIR selection criterion may suggest that the faint end of
the quasar UV luminosity function is steeper at these redshifts than
previously thought \citep{2007ApJ...654..731H, 2012ApJ...746..125H}.
Using far-UV spectral slopes from composite spectra of low-redshift
quasars, and assuming a Lyman continuum escape fraction of 100\%,
\citet{2015A&A...578A..83G} argued that AGN brighter than
$M_{1450}=-18$ can potentially produce all of the metagalactic
hydrogen photoionization rate inferred from the \lya forest at
$4<z<6$.  Second, \citet{2015MNRAS.447.3402B} reported a large scatter
in the \lya opacity between different sightlines close to redshift
$z=6$. \cite{2015MNRAS.453.2943C} showed that these opacity
fluctuations extend to substantially larger scales ($\gtrsim 50\,
h^{-1}$cMpc) than expected in reionization histories dominated by
low-luminosity galaxies (see also \citealt{2016MNRAS.460.1328D}).
\citet{2017MNRAS.465.3429C} further demonstrated that opacity
fluctuations on such large scales arise naturally if there is a
significant contribution ($\gtrsim 50\%$) of AGN to the ionising
emissivity at the redshift of the observed opacity fluctuations ($z
\sim 5.5$--$6$) as would be expected for a QSO luminosity that is
consistent with the measurements of \citet{2015A&A...578A..83G}.
Third, measurements of the Lyman continuum escape fraction from
high-redshift galaxies are still elusive.  Although high-redshift
galaxies as faint as rest-frame UV magnitude $M_\mathrm{UV}=-12.5$
($L\sim 10^{-3}L^*$) at $z=6$ \citep{2017ApJ...835..113L} and
redshifts as high as $z=11.1$ \citep{2016ApJ...819..129O} have now
been reported, the escape of Lyman continuum photons has been detected
in only a small number of comparatively bright ($L>0.5L^*$)
low-redshift ($z < 4$) galaxies.  In these galaxies, the escape
fraction is typically found to be 2--20\% \citep{2010ApJ...725.1011V,
  2011ApJ...736...41B, 2015ApJ...804...17S, 2015ApJ...810..107M,
  2016A&A...585A..48G, 2017MNRAS.468..389J, 2017MNRAS.465..316M} but
reionization would require escape fraction of about 20\% in galaxies
down to $M_\mathrm{UV}=-13$ \citep{2016PASA...33...37F,
  2015ApJ...802L..19R, 2016MNRAS.457.4051K}.  Finally, incidence of
high-redshift AGN is also consistent with the shallow bright-end
slopes of the high-redshift ($z\sim 7$) UV luminosity function of
galaxies relative to a Schechter-function representation
\citep{2012MNRAS.426.2772B, 2014MNRAS.440.2810B, 2014ApJ...792...76B,
  2015MNRAS.452.1817B} and the hard spectra of these bright galaxies
\citep{2015MNRAS.450.1846S, 2015MNRAS.454.1393S, 2017MNRAS.464..469S}.

It is thus pertinent to ask what a significant contribution of AGN to
the ionising emissivity during reionization implies for the search for
the 21~cm signal from the epoch of reionization.  In this paper, we
therefore present predictions for the 21~cm power spectrum from
redshifts $z=7$--$10$ in models of reionization in which the
hydrogen-ionizing emissivity is dominated by AGN and compare them to
galaxy-dominated models.  We use the excursion set method to derive
the large-scale ionization field in high-dynamic-range cosmological
simulations, using the calibration scheme developed by
\citet{2015MNRAS.452..261C} to incorporate 21~cm signal from
self-shielded high density regions \citep{2016MNRAS.tmp.1278K}.  AGN
are modelled by placing black holes in haloes by assuming the $z=0$
M$_\mathrm{bh}$--$\sigma$ relation between the black hole mass and the
bulge stellar velocity dispersion \citep{2012MNRAS.422.1306K,
  2002ApJ...578...90F}.  We describe our simulations and the AGN model
in Section~\ref{sec:model}.  Section~\ref{sec:results} presents our
predictions for the 21~cm signal and its observability in ongoing and
future experiments.  We discuss the case for and against reionization
by AGN in Section~\ref{sec:foragainst}, and end by summarising our
results in Section~\ref{sec:discuss}.  Our $\Lambda$CDM cosmological
model has $\Omega_\mathrm{b}=0.0482$, $\Omega_\mathrm{m}=0.308$,
$\Omega_\Lambda=0.692$, $h=0.678$, $n=0.961$, $\sigma_8=0.829$, and
$Y_\mathrm{He}=0.24$ \citep{2014A&A...571A..16P}.

\section{Models of AGN-dominated reionization}
\label{sec:model}

Our 21 cm predictions are based on cosmological hydrodynamical
simulations that are part of the Sherwood simulation suite
(nottingham.ac.uk/astronomy/sherwood;
\citealt{2017MNRAS.464..897B}). Sources of ionizing radiation are
placed in haloes identified in the simulation, and an ionization field
is obtained using the well-known excursion set approach.  This
ionization field is then calibrated to a given reionization history,
while accounting for residual neutral gas in high-density areas within
ionized regions.  The reionization histories used for calibration are
chosen carefully such that they are consistent with \lya and CMB data
as described in \citet{2015MNRAS.452..261C}.  In this manner, our
models self-consistently predict, at high resolution, the large-scale
distribution of neutral hydrogen for reionization histories consistent
with constraints during the late stages of reionization.

\citet{2016MNRAS.tmp.1278K} provide more details of our implementation
of the excursion set method of deriving the large-scale ionization
field and its subsequent calibration to \lya and CMB data.  We
recapitulate an outline of the method here to mention important
parameter values and set up notation.  We obtain the gas density field
from the underlying cosmological simulation by projecting the relevant
particles onto a grid using the cloud-in-cell (CIC) scheme.  From the
gas density field, we derive the ionization field corresponding to a
distribution of sources with specific ionizing emissivities.  Denoting
the total number of ionizing photons produced by a halo of mass $M$ as
$N_\gamma(M)$, a grid cell at position $\mathbf{x}$ is ionized if the
condition
\begin{equation}
  \zeta_\mathrm{eff}f(\mathbf{x},R)\geq 1
  \label{eqn:exset}
\end{equation}
is satisfied in a spherical region centred on the cell for some radius
$R$ \citep{2004ApJ...613....1F, 2009MNRAS.394..960C,
  2011MNRAS.411..955M}.  Here,
\begin{equation}
  f\propto \rho_m(R)^{-1}\int_{M_\mathrm{min}}^\infty dM\left.
  \frac{dN}{dM}\right\vert_{R}N_\gamma(M),
\end{equation}
where $\rho_m(R)$ is the average matter density and $dN/dM|_R$ is
the halo mass function in the sphere of radius $R$ and
$M_\mathrm{min}$ is the minimum mass of halos that emit Lyman
continuum photons.  The quantity $f$ is proportional to the collapsed
fraction $f_\mathrm{coll}$ into haloes of mass $M>M_\mathrm{min}$ if
$N_\gamma(M)\propto M$.  The parameter $\zeta_\mathrm{eff}$ here is
the effective ionizing efficiency, which corresponds to the number of
photons in the IGM per hydrogen atom in stars, compensated for the
number of hydrogen recombinations in the IGM.  It is the only
parameter that determines the large scale ionization field in this
approach.  Cells that do not satisfy the criterion in
Equation~(\ref{eqn:exset}) are neutral.  We denote the ionized volume
fraction in a cell $i$ as $Q_i$.  The total volume-weighted ionized
fraction is then $Q_V\equiv \sum_iQ_i/n_\mathrm{cell}$, where
$n_\mathrm{cell}$ is the total number of grid cells.

\begin{figure*}
  \begin{center}
    \includegraphics[width=\textwidth]{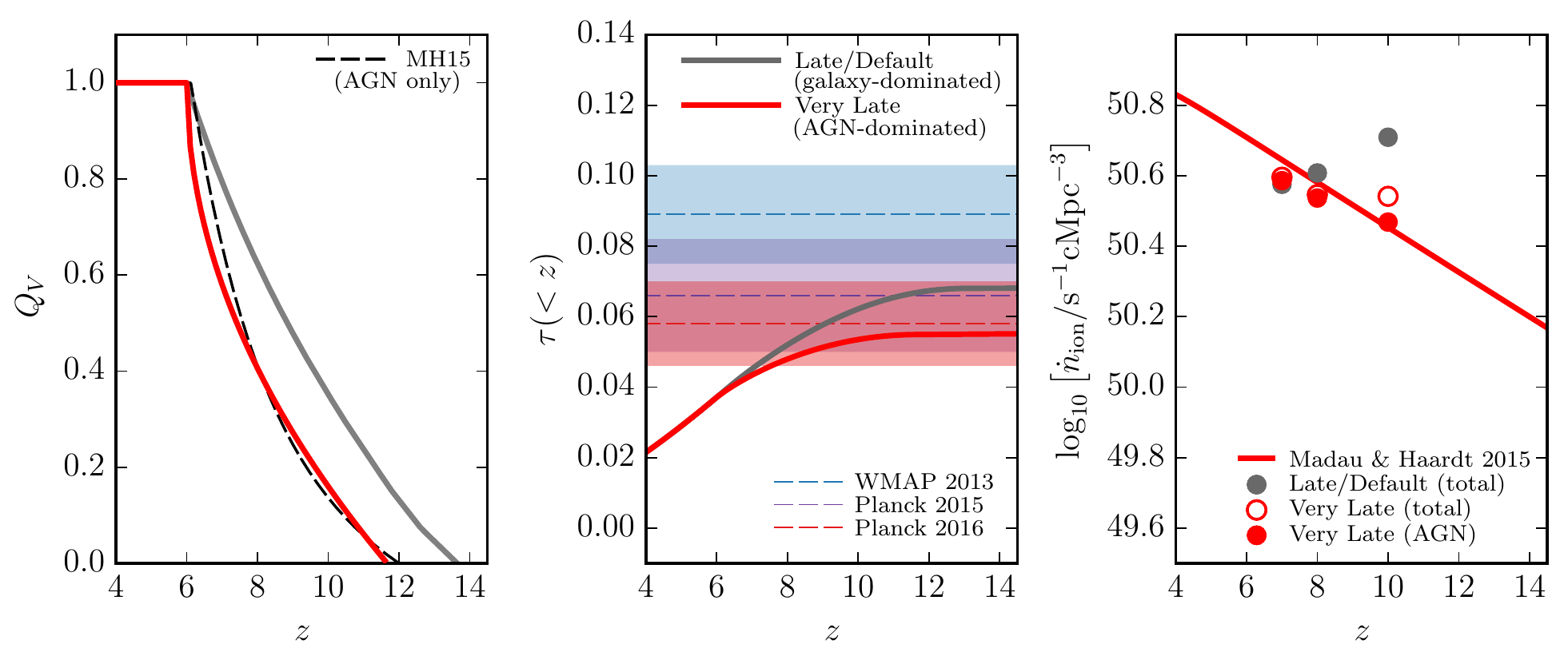}
  \end{center}
  \caption{Evolution of the volume-weighted ionization fraction $Q_V$
    (left panel), the electron cattering optical depth $\tau$ (middle
    panel), and the hydrogen-ionizing emissivity $\dot n_\mathrm{ion}$
    (right panel) in the AGN-dominated ``Very Late'' reionization
    model considered in this paper and the galaxies-dominated
    ``Late/Default'' model of \citet{2016MNRAS.tmp.1278K}.  The
    AGN-dominated model is shown by red curves in left and middle
    panels, and red points and open circles in the right panel.  The
    galaxy-dominated model is shown by the grey curves and points.
    The dashed black curve in the left panel shows the model of
    \citet{2015ApJ...813L...8M}, which includes emission only from
    AGN; the red curve in the right panel shows the corresponding
    emissivity.  In the right panel, red filled circles denote
    ionizing emissivity from AGN in our model; red open circles refer
    to the total ionizing emissivity, which also includes contribution
    from star-forming galaxies.}
  \label{fig:ndotion}
\end{figure*}

We calibrate the large-scale ionization field obtained by the above
procedure to a chosen reionization history, incorporating
inhomogeneities within ionized regions, using the method developed by
\citet{2015MNRAS.452..261C}.  We begin by fixing a reionization model,
which is specified by the redshift evolution of the volume-weighted
ionization fraction $Q_V$.  Our simulated ionization field is
calibrated to the given reionization model in two steps.  In the first
step, the effective ionization parameter $\zeta_\mathrm{eff}$ is tuned
to get the volume-weighted ionization fraction predicted by the
reionization history at the corresponding redshift.  In the second step,
we obtain the photoionization rate distribution within the ionized
regions by solving the globally averaged radiative transfer equation
\begin{equation}
  \frac{dQ_V}{dt}=\frac{\dot n_\mathrm{ion}}{n_\mathrm{H}}-\frac{Q_V}{t_\mathrm{rec}}
  \label{eqn:reion}
\end{equation}
for the photoionization rate $\Gamma_\mathrm{HI}$.  Here, $\dot
n_\mathrm{ion}$ is the average comoving photon emissivity,
$n_\mathrm{H}$ is the average hydrogen density, and $t_\mathrm{rec}$
is the recombination time-scale.  We implement self-shielding in
ionized regions using the fitting function obtained by
\citet{2013MNRAS.430.2427R} from radiative transfer
simulations.\footnote{This self-shielding is insensitive to the
    presence of hard ionizing photons, such as those from AGN, due to
    diminished ionization cross-section \citep{2013MNRAS.430.2427R}.}
This creates cells with excess neutral hydrogen fraction, thereby
reducing the mean free path of Lyman continuum photons.  The mean free
path $\lambda_\mathrm{mfp}$ enters Equation~(\ref{eqn:reion}) via
$\dot n_\mathrm{ion}$, which is related to the photoionization rate by
\citep{2012MNRAS.423..862K, 2013MNRAS.436.1023B}
\begin{equation}
  \dot n_\mathrm{ion} = \frac{\Gamma_\mathrm{HI}Q_V}{(1+z)^2
    \sigma_\mathrm{H}\lambda_\mathrm{mfp}}\left(
  \frac{\alpha_b + 3}{\alpha_s}\right),
  \label{eqn:term1}
\end{equation}
where $\sigma_\mathrm{H}$ is the hydrogen photoionization
cross-section, $\alpha_s$ is the spectral index of the ionizing
sources at $\lambda<912$~{\AA} and $\alpha_b$ is the spectral index of
the ionizing ``background'' within ionized regions.

The Sherwood simulation suite has been run using the energy- and
entropy-conserving TreePM smoothed particle hydrodynamical (SPH) code
\textsc{p-gadget-3}, which is an updated version of the
\textsc{gadget-2} code \citep{2001NewA....6...79S,
  2005MNRAS.364.1105S}.  Our base simulation was performed in a
periodic cube of length 160 $h^{-1}$cMpc on a side.  The number of gas
and dark matter particles were both initially $2048^3$.  This
corresponds to a dark matter particle mass of
$M_\mathrm{dm}=3.44\times 10^7$ $h^{-1}$M$_\odot$ and gas particle
mass of $M_\mathrm{gas}=6.38\times 10^6$ $h^{-1}$M$_\odot$.  In the
redshift range relevant to this paper, we use snapshots of the
particle positions at $z=10, 8,$ and $7$.  Haloes are identified using
the friends-of-friends algorithm.  At $z=7$, the minimum halo mass in
our simulation is $2.3\times 10^{8}$~$h^{-1}$M$_\odot$; the maximum
halo mass is $3.1\times 10^{12}$~$h^{-1}$M$_\odot$.

To model ionizing emission by AGN, we assume that in high-mass haloes
that host luminous AGN, the total number of photons $N_\gamma$ is
proportional to the black hole mass $M_{\rm bh}$.  In order to
estimate the mass of black holes in these haloes, we follow the
approach of \citet{2012MNRAS.422.1306K} and employ the $M_{\rm
  bh}$--$\sigma$ relation \citep[cf.][]{2016ApJ...828...96M}.  The
virial velocity (defined as the circular velocity at virial radius)
for a halo of mass $M$ at redshift $z$ is given by
\begin{multline}
  v_c=23.4\,\mathrm{km}\,\mathrm{s}^{-1}
  \left(\frac{M}{10^8h^{-1}M_\odot}\right)^{1/3}\\
  \times\left[\frac{\Omega_m}{\Omega_m^z}
    \frac{\Delta_c}{18\pi^2}\right]^{1/6}
  \left(\frac{1+z}{10}\right)^{1/2},
\end{multline}
where
\begin{equation}
  \Omega_m^z=\frac{\Omega_m(1+z)^3}
        {\Omega_m(1+z)^3+\Omega_\Lambda+\Omega_k(1+z)^2},
\end{equation}
and $\Delta_c$ is the overdensity of the halo relative to the critical
density, given by
\begin{equation}
\Delta_c=18\pi^2+82d-39d^2,
\end{equation}
where $d=\Omega_m^z-1$ \citep{2001PhR...349..125B}.  Further, we
equate the halo virial velocity with the circular velocity $v_c$ of
its constituent spheroid and obtain the velocity dispersion of the
spheroid using the relation \citep{2002ApJ...578...90F}
\begin{equation}
  v_c\approx 314\left[\frac{\sigma}{208\,\mathrm{km}\,
      \mathrm{s}^{-1}}\right]^{0.84}\mathrm{km}\,\mathrm{s}^{-1}.
\label{eqn:sphsigma}
\end{equation}
Equation~(\ref{eqn:sphsigma}) combined with the measured $M_{\rm
  bh}$--$\sigma$ relation at redshift $z=0$
\citep{2002ApJ...574..740T}
\begin{equation}
  \frac{\sigma}{208\,\mathrm{km}\,\mathrm{s}^{-1}}\approx
  \left[\frac{M_\mathrm{bh}}{1.56\times 10^8\mathrm{M}_\odot}\right]^{1/4.02},
\label{eqn:msigma}
\end{equation}
gives 
\begin{equation}
  \frac{M_\mathrm{bh}}{10^{8}\mathrm{M}_\odot}=0.12
  \left(\frac{M_\mathrm{halo}}{10^{12}\mathrm{M}_\odot}\right)^{1.6}
  \left[\frac{\Omega_m}{\Omega_m^z}
    \frac{\Delta_c}{18\pi^2}\right]^{0.8}(1+z)^{2.4}.
\label{eqn:bhmass}
\end{equation}

\begin{figure*}
  \begin{center}
    \includegraphics*[width=\textwidth]{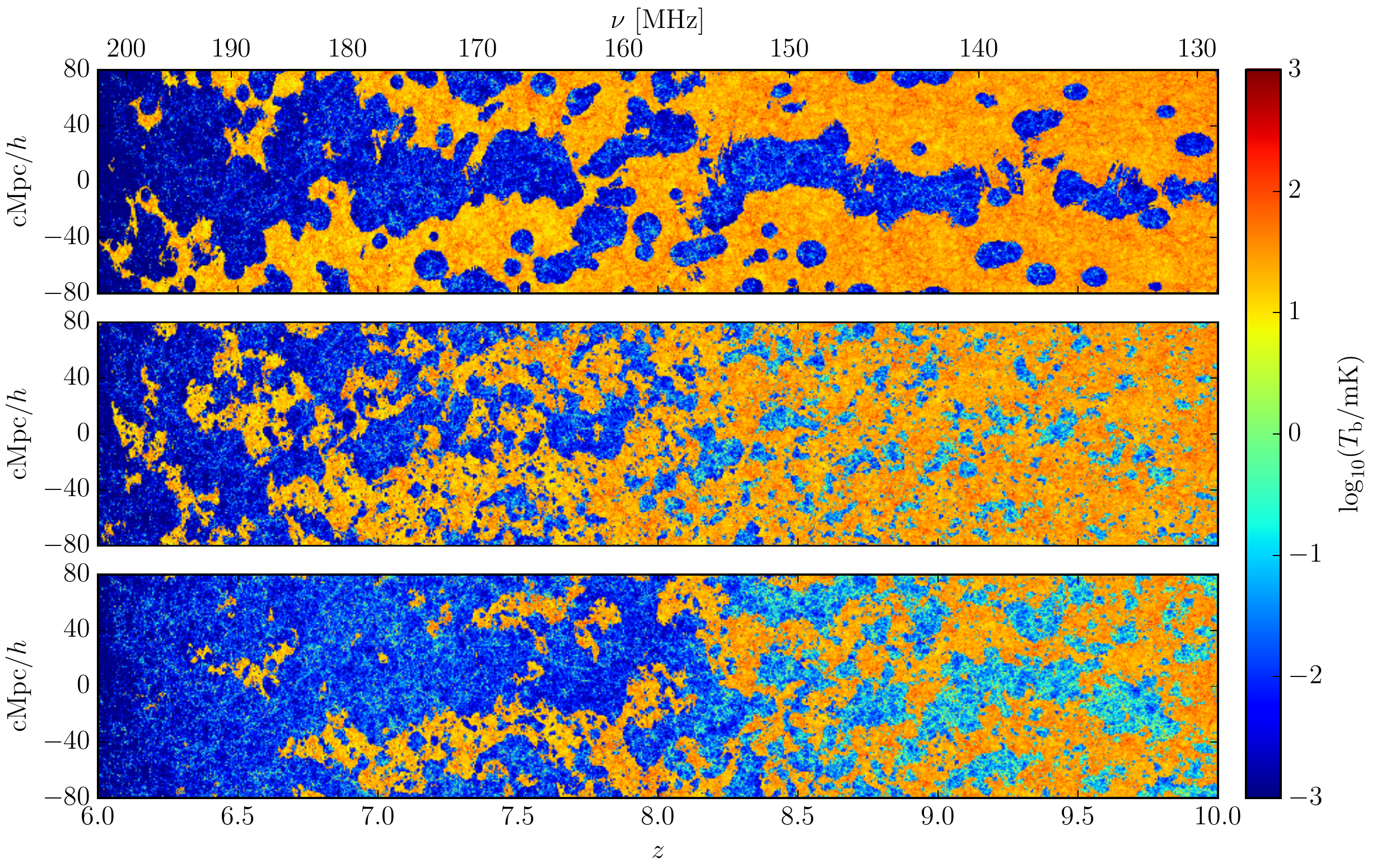}
  \end{center}
  \caption{Evolution of the 21~cm brightness temperature distribution
    from redshift $z=10$ to $6$ in the AGN-dominated Very Late model
    (top panel) introduced in this paper, the galaxies-dominated Very
    Late model (middle panel) from \citet{2016MNRAS.tmp.1278K}, and
    the galaxies-dominated Late/Default model (bottom panel) from
    also \citet{2016MNRAS.tmp.1278K}.}
\label{fig:lightcones}
\end{figure*}

We assume that haloes with mass below a threshold mass $M_q$ have
$N_\gamma(M)=N_\gamma^\mathrm{gal}(M)\propto M_\mathrm{halo}$.
Ionizing photons from these low-mass haloes are sourced by star
formation.\footnote{While AGN have harder spectra than
    star-forming galaxies, the effect of harder photons on the
    structure of the hydrogen ionization fronts is small
    \citep{2008MNRAS.384.1080T, 2015MNRAS.447.1806G,
      2016arXiv160707744K}.  The excursion set method therefore
    remains applicable.} On the other hand, high-mass haloes with
mass greater than the threshold $M_q$ have
$N_\gamma(M)=N_\gamma^\mathrm{agn}(M)\propto M_\mathrm{bh}$ where
$M_\mathrm{bh}$ is given by Equation~(\ref{eqn:bhmass}).  These
high-mass haloes produce ionizing photons due to AGN.  The ratio
\begin{equation}
  r\equiv\frac{\int_{M_q}^{M_\mathrm{max}}dM N_\gamma^\mathrm{agn}(M)\, dN/dM}
  {\int_{M_\mathrm{min}}^{M_q}dMN_\gamma^\mathrm{gal}(M)\, dN/dM},
  \label{eqn:r}
\end{equation}
quantifies the relative photon contribution of AGN and galaxies.  Our
AGN models are thus described by two parameters $r$ and $M_q$.
(Appendix~\ref{sec:ngamma} gives further details on our AGN ionizing
emissivity model.)

In our fiducial AGN-dominated model, we fix the value of the threshold
mass $M_q$ to that corresponding to a circular velocity of $v_c=175$
km$/$s.  (We will discuss the effect on our results of changing this
threshold to $v_c=150$ km$/$s and $v_c=200$ km$/$s below.)  At lower
circular velocities, cold gas mass available to grow supermassive
black holes can decrease rapidly due to an increasing effect of
supernova feedback \citep[e.g.,][]{2000MNRAS.311..576K, 2002MNRAS.336L..61H,
  2012MNRAS.419..771B}.  This is reflected in a drop in the black hole
mass function for black hole masses smaller than $M_\mathrm{bh}\sim
10^7~\mathrm{M}_\odot$, particularly for $z>1$
\citep{2008MNRAS.388.1011M, 2012AdAst2012E...7K}.  With $M_q$ fixed, a
desired total AGN emissivity is achieved in the model by setting the
value of the parameter $r$.  We calibrate the AGN emissivity evolution
to values close to the fit by \citet{2015ApJ...813L...8M} to the
integrated 1~Ry emissivity from AGN down to UV luminosities of $0.01
L_*$.  This emissivity evolution is shown by the red curve in the
right panel of Figure~\ref{fig:ndotion}.  In this panel, red filled
circles denote ionizing emissivity from AGN in our model; red open
circles refer to the total ionizing emissivity, which also includes
contribution from star-forming galaxies.  The ionizing emissivity of
AGN in our model closely matches that from the model of
\citet{2015ApJ...813L...8M}.  We also have some contribution to $\dot
n_\mathrm{ion}$ from star-forming galaxies in our model, particularly
at $z=10$, as seen from the red open circles in
Figure~\ref{fig:ndotion}.  For comparison, the grey points in
Figure~\ref{fig:ndotion} show the photon emissivity in the
galaxy-dominated ``Late/Default'' model of
\citet{2016MNRAS.tmp.1278K}.

\begin{figure*}
  \begin{center}
    \includegraphics[width=\textwidth]{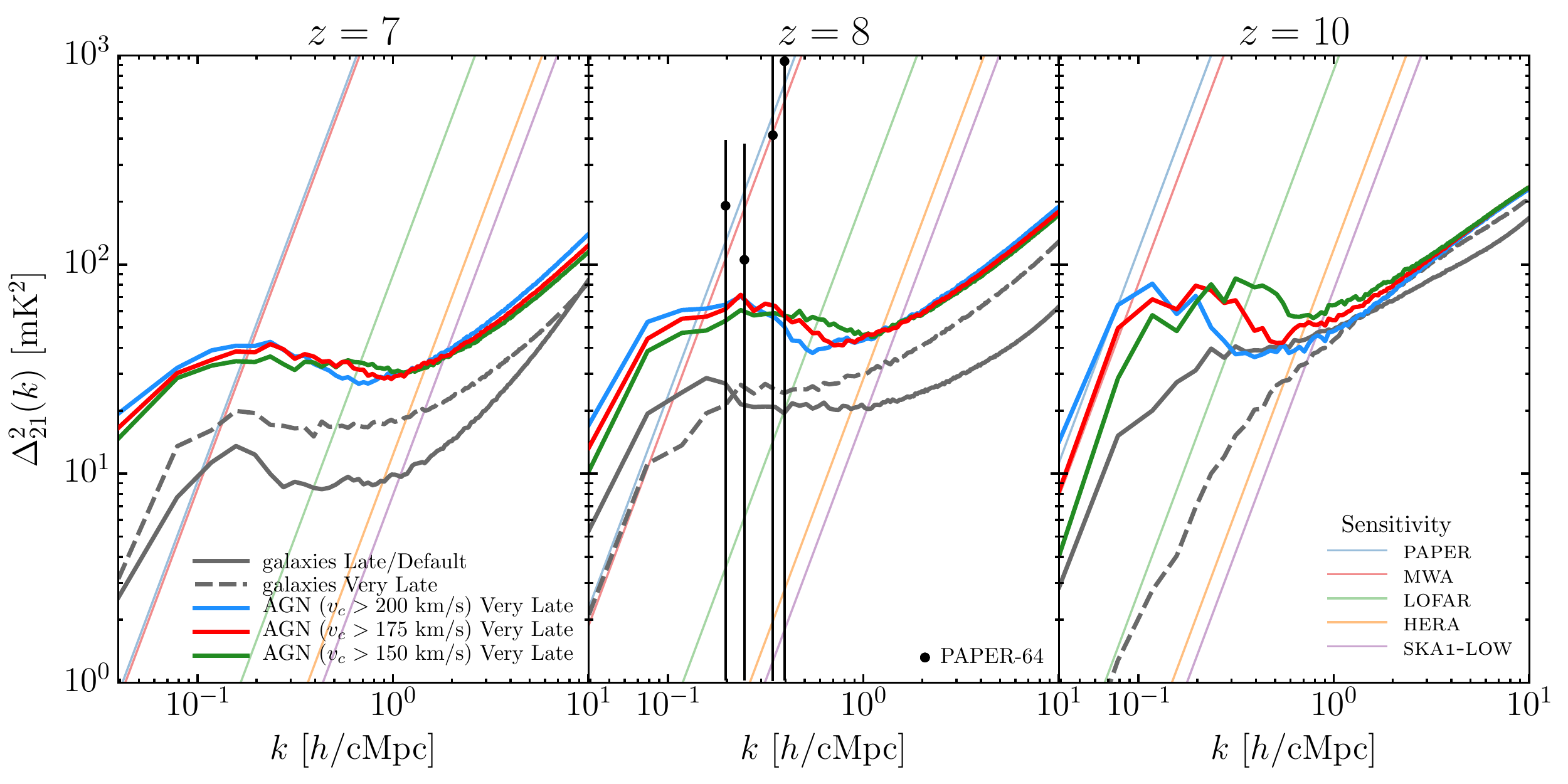}
  \end{center}
  \caption{Red curves show the 21~cm power spectra from our fiducial
    AGN-dominated model, which assumes that AGN are hosted by haloes
    with circular velocities greater than $v_c=175$~km/s.  Green and
    blue curves show the power spectra in models where this threshold
    is changed to 150~km/s and 200~km/s, respectively.  All three of
    these models are calibrated to the Very Late reionization history.
    Power spectra from the galaxy-dominated Late/Default model of
    \citet{2016MNRAS.tmp.1278K} are shown by the solid grey curves.
    Dashed grey curves show power spectra from a galaxy-dominated Very
    Late model, also from \citet{2016MNRAS.tmp.1278K}.  Thin coloured
    diagonal lines indicate experimental sensitivites.  Data points
    are the measurements from the 64-element deployment of PAPER at
    $z=8.4$ \citep{2015ApJ...809...61A}.  The average ionization
    fraction in the Very Late model is $Q_V=0.16$ at $z=10$, $0.41$ at
    $z=8$, and $0.58$ at $z=7$.  At these redshifts, the average
    ionization fraction in the Late/Default model is, respectively,
    $Q_V=0.37$, $0.65$, and $0.82$.}
  \label{fig:result3}
\end{figure*}

Having chosen a source model, we now need to choose a suitable
reionization history to calibrate our simulation.  As discussed above
in relation to Equation~(\ref{eqn:reion}), this calibration will
provide us with the photoionization rate and ionized hydrogen fraction
throughout our simulation box.  The AGN-dominated ionizing emissivity
evolution considered by \citet{2015ApJ...813L...8M} gives rise to a
reionization history that is very close to the ``Very Late''
reionization history as described by \citet{2016MNRAS.tmp.1278K}.  For
ease of comparison we thus choose this reionization history to
calibrate our simulation.  The red curve in the left panel of
Figure~\ref{fig:ndotion} shows the evolution of the volume-weighted
ionized fraction $Q_V$ in this Very Late model.  For comparison, the
grey curve in this panel shows the evolution of $Q_V$ in the
``Late/Default'' model of \citet{2016MNRAS.tmp.1278K}.  In the Very
Late model, ionized regions overlap and the universe is completely
reionized at $z = 6$, similar to the Late/Default model, but $Q_V$
evolves more rapidly at $z > 6$ \citep{2016MNRAS.tmp.1278K}.  This
model agrees reasonably well with the background photoionization rate
determined from the Ly$\alpha$ forest at $z<6$
\citep{2009ApJ...703.1416F, 2013MNRAS.436.1023B} and from quasar
proximity zones at $z\sim 6$ \citep{2011MNRAS.412.1926W,
  2011MNRAS.412.2543C}, albeit with notable differences
\citep{2015MNRAS.450.4081P, 2015MNRAS.453.2943C}.  The value of the
electron scattering optical depth to the last scattering surface in
this model is $\tau = 0.055$, in good agreement with the most recent
Planck measurement ($\tau=0.058\pm 0.012$;
\citealt{2016A&A...596A.108P}), as seen in the middle panel of
Figure~\ref{fig:ndotion}.

\section{Results: 21~cm signal}
\label{sec:results}

The top panel of Figure~\ref{fig:lightcones} shows the evolution of
the 21~cm brightness temperature from redshift $z=10$ to $6$ in our
fiducial $v_c>175$~km/s AGN-dominated model.  The 21~cm brightness
temperature is approximated as
\begin{equation}
  T_b(\mathbf{x})=\overline T_b x_\mathrm{HI}(\mathbf{x})\Delta(\mathbf{x}),
  \label{eqn:tb}
\end{equation}
where the mean temperature $\overline T_b\approx 22 \mathrm{mK}
[(1+z)/7]^{1/2}$ \citep{2009MNRAS.394..960C}.  The above relation
neglects the impact of redshift space distortions due to peculiar
velocities, and possible fluctuations in the spin temperature, i.e.,
it implicitly assumes that the spin temperature is much greater than
the CMB temperature and that the Ly$\alpha$ coupling is sufficiently
complete throughout the IGM.  This is a good approximation in the
redshift range considered here, when the global ionized fraction is
greater than a few per cent \citep{2012RPPh...75h6901P,
  2014MNRAS.443.2843M, 2015MNRAS.447.1806G}.  For comparison, the
middle panel of Figure~\ref{fig:lightcones} shows the evolution of the
21~cm brightness in the galaxies-dominated Very Late model considered
by \citet{2016MNRAS.tmp.1278K}.  The reionization history of this
model is identical to that of our AGN-dominated model, so the
differences in the brightness distribution between the top and middle
panels of Figure~\ref{fig:lightcones} arise solely due to differences
in the source model.  The AGN-dominated model has fewer, larger, and
more clustered ionized regions than the galaxies-dominated model
(cf.\ \citealt{2007MNRAS.377.1043M}).  Figure~\ref{fig:lightcones}
also shows the galaxies-dominated Late/Default model of
\citet{2016MNRAS.tmp.1278K} in the bottom panel.  The source model as
well as the reionization history are now different from our
AGN-dominated model.  This is reflected in a strikingly different
morphology of 21-cm-bright regions.

We derive the power spectrum of the 21~cm fluctuations in our model as
\begin{equation}
  \Delta_{21}^2(k) = \frac{k^3}{2\pi^2}\cdot
  \frac{\langle\widetilde{T_b}^2(k)\rangle}{V_\mathrm{box}},
\end{equation}
where $\widetilde{T_b}(k)$ is the Fourier transform of the brightness
temperature defined in Equation~(\ref{eqn:tb}), the average is over
the simulation box, and $V_\mathrm{box}=(160~\mathrm{cMpc}/h)^3$ is
the comoving box volume.  Figure~\ref{fig:result3} shows our main
results at redshifts $z=7$, $8$, and $10$ in its left, middle, and
right panels, respectively.  The red curve in all panels shows the
21~cm power spectrum in our fiducial AGN-dominated model, in which AGN
are hosted by haloes with $v_c>175$~km/s.  The power spectrum is
characterised by a bump at large scales and an increase towards the
smallest scales.  At redshifts $z=7$--$10$ shown, the bump occurs at
$k\sim 0.2$ cMpc$^{-1}h$ and has an amplitude of approximately
$\Delta_{21}^2\sim 40$--$70$~mK$^2$.  This is significantly higher
than in the galaxies-dominated models.  (Note that $k=0.2$
cMpc$^{-1}h$ corresponds to a length scale of 30~$h^{-1}$cMpc, which
is well-sampled in our simulation cube, which is 160~$h^{-1}$cMpc on
each side.)  We can compare the large-scale power in our AGN-dominated
model with that in the galaxies-dominated Very Late model in
Figure~\ref{fig:result3}, in which the galaxies-dominated model is
shown by the dashed grey curves.  The large-scale power in the
AGN-dominated model is larger than that in the galaxies-dominated
model by factor of 2 at $z=7$ and a factor of 10 at $z=10$.  As we
will see below, this enhancement is due to the enhanced size and
clustering of ionized regions, which is also visually apparent in
Figure~\ref{fig:lightcones}.  The large-scale 21~cm power in our
fiducial AGN-dominated model is also higher than the large-scale power
in the Late/Default model of \citet{2016MNRAS.tmp.1278K}.  Power
spectra from the latter model are shown by the solid grey curves in
Figure~\ref{fig:result3}.  The enhancement factor here is about 3 at
$z=7$ and 2 at $z=10$.  At redshifts $z=8$ and $10$, the Late/Default
model has higher power than the galaxies-dominated Very Late model at
large scales, because of the higher $Q_V$, which translates to larger
bubble size, as is evident from Figure~\ref{fig:lightcones}.

In Figure~\ref{fig:result3} we also show the effect of changing the
circular velocity threshold for AGN-hosting haloes.  The red curve in
this figure shows the 21~cm power spectrum in our fiducial
AGN-dominated model, when AGN are hosted by haloes with
$v_c>175$~km/s.  The blue and green curves show the power spectra when
the circular velocity threshold is changed to 200~km/s and 150~km/s,
respectively.  The main effect of this change on the large-scale 21~cm
power is to shift the position of the bump.  As all three models are
calibrated to the same Very Late reionization history, they have
identical ionization fraction $Q_V$ at each redshift.  Therefore, when
the circular velocity threshold is reduced, the number of AGN-hosting
haloes increases and consequently, in order to hold $Q_V$ fixed, the
size of individual ionized regions decreases.  This moves the bump in
the 21~cm power spectrum to smaller scales
(cf.\ \citealt{2012MNRAS.423.2222I}).

The thin diagonal lines in each panel of Figure~\ref{fig:result3} show
sensitivities set by thermal noise for five ongoing and upcoming 21~cm
experiments: the Precision Array for Probing the Epoch of Reionization
(PAPER; \citealt{2014ApJ...788..106P}), Murchison Widefield Array
(MWA; \citealt{2013PASA...30...31B, 2013PASA...30....7T}), Low
Frequency Array (LOFAR; \citealt{2013A&A...556A...2V};
\citealt{2014ApJ...782...66P}), Hydrogen Epoch of Reionization Array
(HERA; \citealt{2014ApJ...782...66P}), and the low frequency
instrument from Phase 1 of the Square Kilometre Array (SKA1-LOW).  We
consider 1000~hr of observations and use experimental parameters
identical to those considered by \citet{2016MNRAS.tmp.1278K}.  Note
that the sample variance from the limited number of $k$-modes measured
in the survey volume also limits the sensitivity of the experiment.
The sample variance scales as $\Delta^2(k)/\sqrt{N}$ and, due to the
small amplitude of the power spectrum, is smaller ($<1$ mK$^2$) than
the thermal noise at all redshifts for all experiments considered
here.  Also note that we assume perfect foreground subtraction in this
discussion.  Foreground subtraction and calibration residuals will
reduce the experimental sensitivity \citep{2009A&A...500..965B,
  2013ApJ...768L..36P, 2014PhRvD..89b3002D}.  Due to the relatively
smooth dependence of astrophysical foregrounds on frequency, this
reduction in sensitivity particularly affects small $k$ values.

\begin{figure}
  \begin{center}
    \includegraphics[width=\columnwidth]{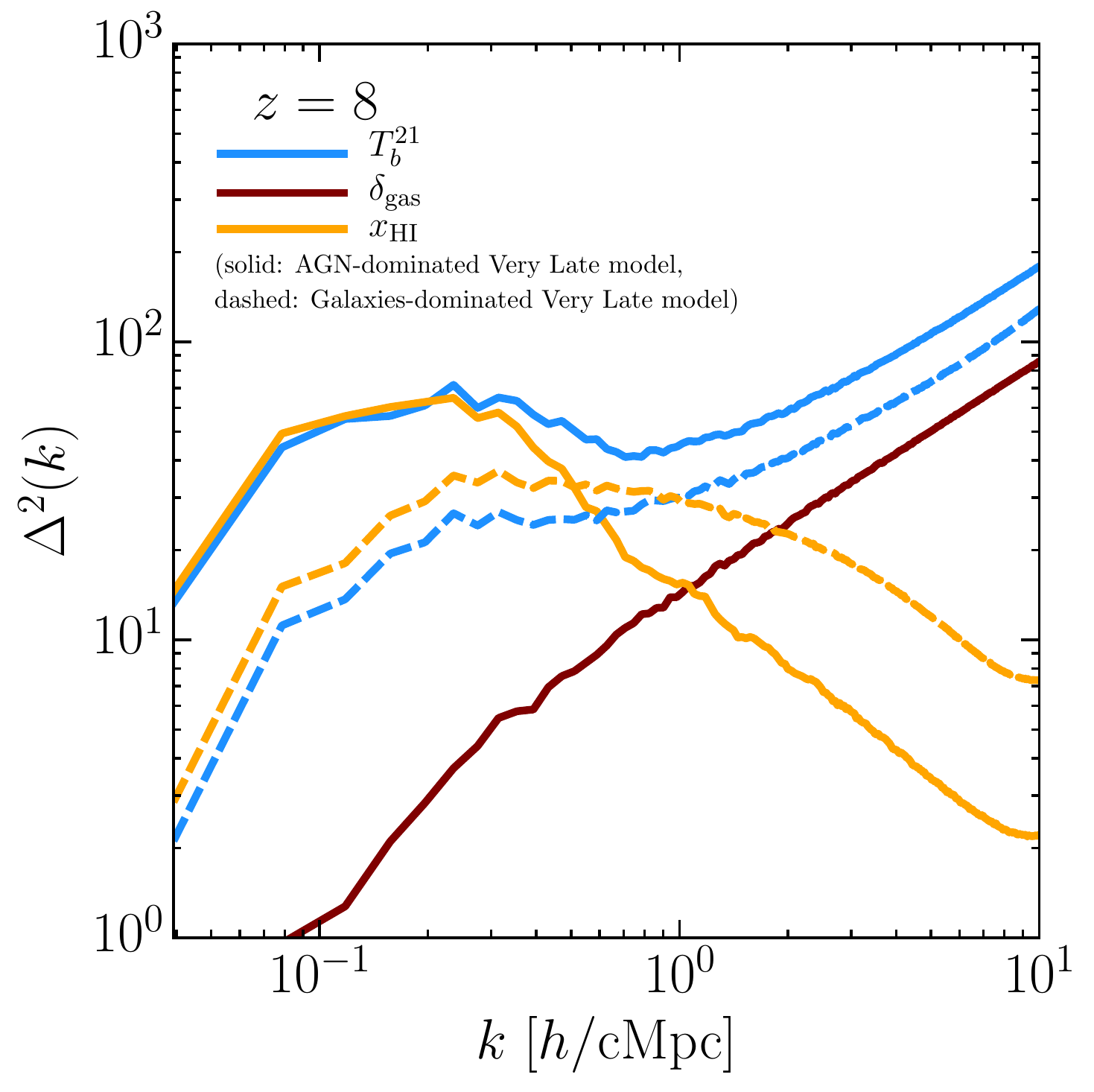}
  \end{center}
  \caption{A decomposition of the 21~cm power spectrum (blue curve)
    predicted in our model into contributions from the gas density
    (brown curve) and the ionization field (orange curve) at $z=8$.
    Solid curves show the AGN-dominated model; dashed curves show the
    galaxies-dominated model.  The gas density power spectrum is
    identical in the two models.  In AGN-dominated models, the
    ionization field is highly clustered.  This explains the
    enhancement in 21~cm power at large scales.}
  \label{fig:ps}
\end{figure}

Due to limited baselines, current and upcoming 21~cm experiments are
only sensitive to large scales.  None of the experiments are sensitive
to 21~cm power at $k\gtrsim 1$ cMpc$^{-1}h$.  SKA1-LOW and HERA have
the highest sensitivities primarily due to large number of antenna
elements.  In the galaxies-dominated Late/Default model, at $z=10$
(129~MHz), the signal to noise ratio (SNR) is $\sim 100$ for these two
experiments at $k\sim 0.1$ cMpc$^{-1}h$.  This is enhanced by a
further factor of $\sim 2$ in the AGN-dominated model.  At $z=7$
(178~MHz), the enhancement is by a factor of $\sim 3$.  LOFAR is
sensitive at scales corresponding to $k\lesssim 0.2$ cMpc$^{-1}h$ at
$z=10$ (129~MHz) and $k\lesssim 0.5$ cMpc$^{-1}h$ at $z=7$ (178~MHz).
At $k\sim 0.1$ cMpc$^{-1}h$, the expected SNR for LOFAR is $\sim 10$
at $z=10$ and $\sim 50$ at $z=7$, for the galaxies-dominated
Late/Default model.  These SNRs are also enhanced by similar factors
as for SKA1-LOW and HERA in AGN-dominated models.  PAPER and MWA are
the least sensitive of the five experiments, due to their relatively
small number of antenna elements.  Figure~\ref{fig:result3} shows that
while galaxies-dominated models predict small SNRs for PAPER and MWA,
the AGN-dominated models do predict SNRs of $\sim 2$ at $k\sim 0.1$
cMpc$^{-1}h$ at redshifts $z=7$ (178~MHz) and $z=8$ (158~MHz).  Also
shown in Figure~\ref{fig:result3} are published measurements from the
64-element deployment of PAPER at $z=8.4$ \citep{2015ApJ...809...61A}.
These are within a factor of $<2$ of the predicted power in our
AGN-dominated model.  Very clearly, the 21~cm signal will be
significantly easier to detect if reionization is AGN-dominated.
Conversely, these could become the first non-trivial models of
reionization to be ruled out by 21~cm experiments, thereby
constraining the contribution of AGN to reionization and thus
complementing infrared surveys.

The enhancement in the large-scale 21~cm power in AGN-dominated
reionization models can be better understood by decomposing the 21~cm power
spectrum into contributions from the ionization field and the
underlying matter density using Equation~(\ref{eqn:tb}).  This yields
\citep{2004ApJ...613....1F, 2005ApJ...630..643M, 2006PhR...433..181F}
\begin{equation}
  \Delta^2_{21}(k)= b_\delta\Delta^2_{\delta}(k)+
  b_x\Delta^2_{x_\mathrm{HI}}(k)+\textrm{cross-correlations}
  \label{eqn:ps_decomp}
\end{equation}
where the first and second terms on the right-hand side are the power
spectra of the matter density and the ionization field, respectively,
and the last term denotes the cross-power spectrum between the
ionization and matter density fields.  The proportionality factors
$b_\delta$ and $b_x$ are independent of $k$.  This decomposition is
shown in Figure~\ref{fig:ps} for our fiducial AGN-dominated model and
the galaxies-dominated Very Late model at $z=8$.  At small scales the
21~cm power spectrum is proportional to the matter power spectrum.  At
large scales, the cross-terms in Equation~\ref{eqn:ps_decomp} are
negative, with a magnitude of about 10\% of the total power.  The
ionization field starts contributing power at large scales, creating a
bump \citep{2006PhR...433..181F}.  This can be understood by writing
the power spectrum of the ionization field in terms of the size
distribution of ionized regions in a halo model approach
\citep{2004ApJ...613....1F, 2005ApJ...630..643M}.  The scale at which
the bump appears depends on the characteristic size of ionized regions
and grows with decreasing redshift.  When the ionization fraction is
small even the large scale power is determined by the matter power
spectrum.  This is the case, for instance, in the galaxy-dominated
Late/Default model at $z=10$ in Figure~\ref{fig:result3}.  However, as
ionized regions grow, the bump moves to successively smaller $k$
values.  This happens with decreasing redshifts, but in our case it
also happens when when we put AGN in successively higher mass haloes,
that is, when we increase the threshold circular velocity of
AGN-hosting haloes, because all of our AGN-dominated models are
calibrated to the same Very Late reionization history.  When we
increase the circular velocity cut-off, the number of AGN-hosting
haloes is reduced and the size of ionized regions around each
AGN-hosting halo increases in order to keep $Q_V$ fixed. This
increases the spatial scale at which the power enhancement occurs.
The amplitude of the peak in the power spectrum at large scales,
however, does not increase arbitrarily with the circular velocity
threshold.  At some point, Poisson fluctuations dominate and the power
approaches that corresponding to white noise.  This is clearly seen in
Figure~\ref{fig:result3} in all models: the peak in large-scale power
is enhanced in the AGN-dominated models relative to the
galaxies-dominated models with the same (Very Late) reionization
history, but when $v_c$ is increased beyond 150~km/s, the peak simply
moves to larger scales without increasing in amplitude.  Thus,
enhanced contribution from high mass haloes with constant total
ionization fraction increases the large scale 21~cm power up to a
limit and then moves the location of the peak to larger and larger
scales.  This large scale peak in the 21~cm power is perhaps the most
important reionization signature for 21~cm experiments
\citep{2006PhR...433..181F}.

\section{The case for and against reionization by AGN}
\label{sec:foragainst}

While interest in early reionization by X-rays from faint AGN
\citep{2004MNRAS.350.1107M, 2005MNRAS.356..596M, 2007MNRAS.374..627S}
was motivated by the large value of Thomson scattering optical depth
measured from the first-year WMAP data
($\tau=0.166^{+0.076}_{-0.071}$; \citealt{2003ApJS..148..175S}), there
are now a number of arguments favouring a significant role of normal
QSOs in reionization, as discussed in Section~\ref{sec:intro}: the
suggestion of a rather steep faint end of the QSO luminosity function
at high redshift by \citet{2015A&A...578A..83G}, large \lya opacity
fluctuations at very large scales in QSO absorption spectra
\citep{2015MNRAS.447.3402B,2015MNRAS.453.2943C, 2016MNRAS.460.1328D},
a lack of convincing detections of the escape of Lyman continuum
photons from faint high-redshift galaxies \citep{2016A&A...585A..48G,
  2015ApJ...810..107M, 2015ApJ...804...17S, 2011ApJ...736...41B,
  2010ApJ...725.1011V, 2016PASA...33...37F, 2015ApJ...802L..19R,
  2016MNRAS.457.4051K}, and finally, the emergence of a shallow bright
end of the high-redshift ($z\gtrsim 7$) galaxy luminosity function
\citep{2012MNRAS.426.2772B, 2014MNRAS.440.2810B, 2014ApJ...792...76B,
  2015MNRAS.452.1817B} with many bright galaxies showing possible
AGN-like spectral signatures \citep{2015MNRAS.450.1846S,
  2015MNRAS.454.1393S, 2017MNRAS.464..469S}.  These observations all
point towards a significant presence of luminous AGN at $z>6$,
suggesting that AGN play a major role in reionization
\citep{2015A&A...578A..83G, 2015ApJ...813L...8M, 2015MNRAS.453.2943C,
  2016MNRAS.457.4051K, 2017MNRAS.465.3429C, 2016arXiv160602719M,
  2016arXiv160706467D}.

On the other hand, however, it has also been argued that AGN-dominated
reionization is in tension with several observations.
\citet{2016arXiv160706467D} considered the effect of AGN-dominated
reionization on the Ly$\alpha$ opacity at $z>5$, He~\textsc{ii}
Ly$\alpha$ opacity at $z\sim 3.1$--$3.3$, and the thermal history of
the IGM.  In agreement with \citet{2015MNRAS.453.2943C}, these authors
found that AGN did provide a plausible explanation for the large
fluctuations in the Ly$\alpha$ opacity at $z>5$.  However, they found
that reionization of He~\textsc{ii} occurs much earlier in these
AGN-dominated models (see also \citealt{2016arXiv160602719M}).  For
instance, in the model of \citet{2015ApJ...813L...8M}, He~\textsc{ii}
reionization is complete at $z=4.5$, compared to $z=3$ in the standard
scenario \citep{2012ApJ...746..125H}.  This early Helium reionization
could result in higher IGM temperatures due to the associated
photoheating.  The temperature of the IGM at mean density is enhanced
in AGN-dominated models by factors of $\sim 2$ relative to the
standard models for $z=3.5$--$5$, in conflict with measurements.  This
inconsistency could be avoided by reducing the escape fraction of 4~Ry
photons in AGN, but it is not clear if this can be achieved while
requiring a 100\% escape fraction of 1~Ry photons in order to explain
the Ly$\alpha$ opacity fluctuations.  Further evidence against
AGN-dominated reionization models has emerged from metal-line
absorbers at $z\sim 6$.  In their cosmological radiation
hydrodynamical simulations, \citet{2016MNRAS.459.2299F} find that the
hard spectral slopes of UV backgrounds in AGN-only reionization models
produce too many C~\textsc{iv} absorption systems relative to
Si~\textsc{iv} and C~\textsc{ii} at $z\sim 6$.  However, these
simulations assume an $L_\nu\propto\nu^{-1.57}$ AGN SED at extreme UV
\citep{2001AJ....122..549V, 2002ApJ...579..500T, 2012ApJ...746..125H}.
This slope is marginally harder than recent measurements
($L_\nu\propto\nu^{-1.7}$) from a stack of $z\sim 2.4$ quasars
\citep{2015MNRAS.449.4204L}.  \citet{2016MNRAS.459.2299F} also find
that the N(Si~\textsc{iv})/N(C~\textsc{iv}) column density ratio
measurements prefer a somewhat harder and more intense $>4$~Ry
background than the standard model of \citet{2012ApJ...746..125H}.
Using a large sample of X-ray-selected quasars in the redshift range
$z=0$--$6$, \citet{2017MNRAS.465.1915R} find that the faint end of the
AGN UV luminosity function at $z\sim 6$ is likely to be much shallower
that that reported by \citet{2015A&A...578A..83G}.  In their analysis,
\citet{2017MNRAS.465.1915R} use an AGN obscuration optical depth
($\log N_\mathrm{H}$) cut-off that reproduces low-redshift AGN UV
luminosity functions and an X-ray-to-optical/UV luminosity ratio
calibrated at redshifts $z=0.05$--$4$ \citep{2010A&A...512A..34L}.
These authors argue that the apparent contradiction with the results
of \citet{2015A&A...578A..83G} could be explained by contamination
from the AGN host galaxies.  It has also been recently argued that the
Lyman continuum escape fraction of AGN might not be 100\% as is
usually assumed \citep{2017MNRAS.465..302M}.  This may further reduce
the contribution of AGN to reionization.

A definitive understanding of the AGN contribution to reionization
will perhaps only emerge with deep large-area surveys to detect faint
and intermediate brightness quasars at high redshifts, such as the
Subaru High-z Exploration of Low-Luminosity Quasars (SHELLQs) project
\citep{2016ApJ...828...26M} and the VISTA Extragalactic Infrared
Legacy Survey (VEILS; \citealt{2017MNRAS.464.1693H}), and later with
the Wide Field Infrared Survey Telescope (WFIRST;
\citealt{2013arXiv1305.5422S}) and Euclid \citep{2011arXiv1110.3193L}.

\section{Conclusions}
\label{sec:discuss}

We have presented predictions of the spatial distribution of the 21~cm
brightness temperature fluctuations from AGN-dominated models of
reionization using high-dynamic-range cosmological hydrodynamical
simulations from the Sherwood simulation suite
\citep{2017MNRAS.464..897B} for reionization histories motivated by
constraints from \lya absorption and emission data as well as CMB data
and based on a physically motivated AGN model.

Our main conclusion is that AGN-dominated reionization histories
increase the large-scale 21~cm power by factors of up to ten.
Conventional models typically predict values of 10--20 mK$^2$ for the
variance of the 21~cm brightness temperature at redshifts $z=7$--$10$
at scales accessible to ongoing and upcoming experiments ($k \lesssim
1$ cMpc$^{-1}h$), but AGN-dominated models can increase this variance
to values close to 100 mK$^2$.  This is because AGN reside in few
highly clustered haloes, which increases the peak of the 21~cm power
spectrum and moves the peak to larger scales.  This bodes well for
experiments that seek to detect this feature, and the predicted signal
is lower than the sensitivity claimed to have been already reached by
ongoing experiments by only a factor of about two or less.

Our models for the reionization history and Lyman continuum emissivity
of AGN suggest that detection by LOFAR (and later HERA and SKA1)
should be in easy reach of their design sensitivity, albeit assuming
optimistic foreground subtraction and calibration residuals.
Conversely, these models could become the first non-trivial hydrogen
reionization scenarios to be ruled out by experiments, thereby
complementing infrared searches for high-$z$ AGN, and constraining the
contribution of AGN to reionization.

\section*{Acknowledgments}

We thank the anonymous referee for a thoughtful referee report and
acknowledge helpul discussions with Jonathan Chardin, Colin DeGraf,
Kristian Finlator, Laura Keating and Dylan Nelson.  Support by ERC
Advanced Grant 320596 `The Emergence of Structure During the Epoch of
Reionization' is gratefully acknowledged.  EP gratefully acknowledges
support by the Kavli Foundation.  We acknowledge PRACE for awarding us
access to the Curie supercomputer, based in France at the Tr\'es Grand
Centre de Calcul (TGCC).  This work used the DiRAC Data Centric system
at Durham University, operated by the Institute for Computational
Cosmology on behalf of the STFC DiRAC HPC Facility
(www.dirac.ac.uk). This equipment was funded by BIS National
E-infrastructure capital grant ST/K00042X/1, STFC capital grants
ST/H008519/1 and ST/K00087X/1, STFC DiRAC Operations grant
ST/K003267/1 and Durham University.  DiRAC is part of the National
E-Infrastructure.  This research was supported by the Munich Institute
for Astro- and Particle Physics (MIAPP) of the DFG cluster of
excellence ``Origin and Structure of the Universe''.

\bibliographystyle{mnras}
\bibliography{refs}

\appendix

\section{AGN ionizing emissivity}
\label{sec:ngamma}

\begin{figure*}
  \begin{center}
    \includegraphics[width=\textwidth]{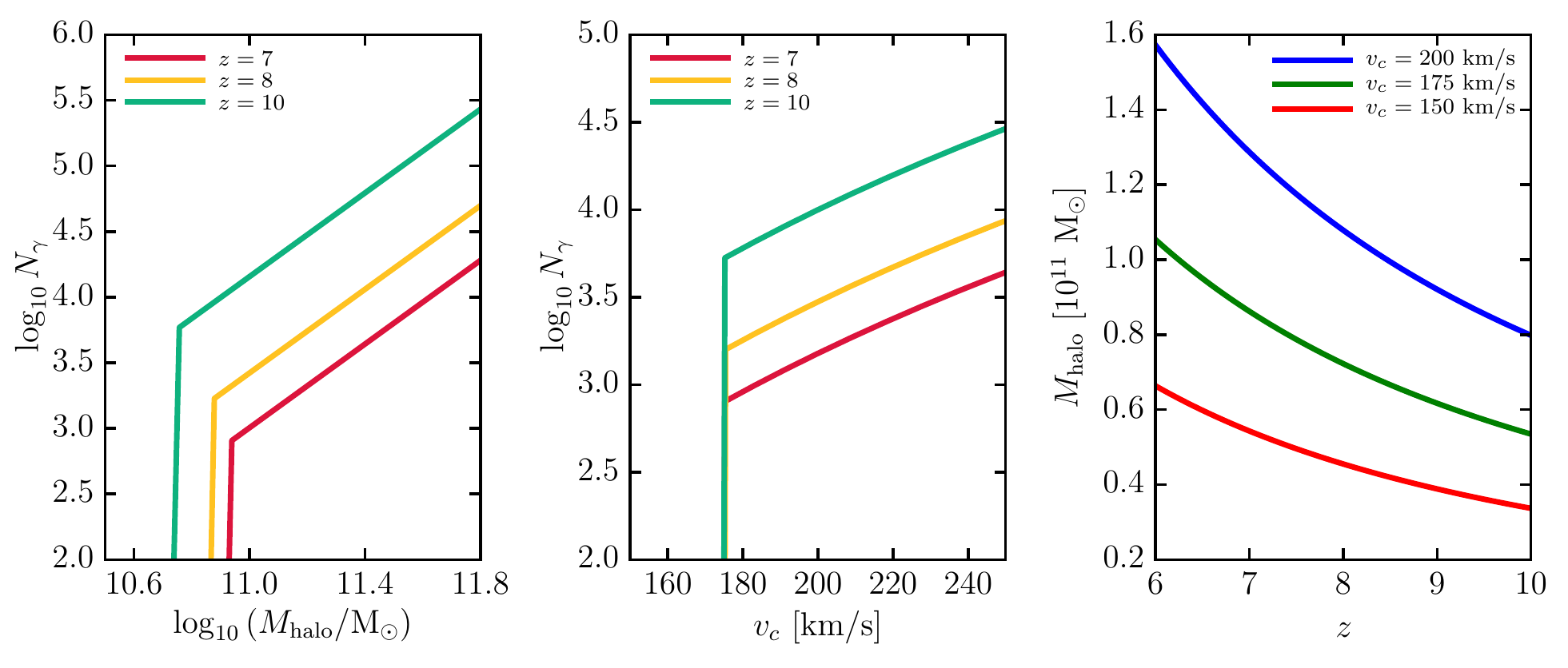}
  \end{center}
  \caption{Our model for the photon contribution $N_\gamma$ for AGN at
    various redshifts as a function of halo mass (left panel) and
    circular velocity (middle panel).  The right panel shows the
    evolution of halo mass corresponding to the three circular
    velocity thresholds considered in this paper.  Our chosen
    reionization model dictates the amplitude of the
    $N_\gamma$--$M_\mathrm{halo}$ relation, while its slope is
    governed by the black hole mass model of
    Equation~(\ref{eqn:bhmass}).}
  \label{fig:ngamma}
\end{figure*}

In our simulations, AGN are implemented according to the procedure
described in Section~\ref{sec:model}.  In this work we assume that
haloes with mass below a threshold mass $M_q$ have
$N_\gamma(M)=N_\gamma^\mathrm{gal}(M)\propto M_\mathrm{halo}$.
Ionizing photons from these low-mass haloes are sourced by star
formation.  On the other hand, high-mass haloes with mass greater than
the threshold $M_q$ have $N_\gamma(M)=N_\gamma^\mathrm{agn}(M)\propto
M_\mathrm{bh}$ where $M_\mathrm{bh}$ is given by
Equation~(\ref{eqn:bhmass}).  These high-mass haloes produce ionizing
photons due to AGN.  The ratio $r$, defined in Equation~(\ref{eqn:r}),
quantifies the relative photon contribution of AGN and galaxies.  Our
AGN models are thus described by two parameters $r$ and $M_q$.  We fix
the value of the threshold mass $M_q$ to that corresponding to a
circular velocity of $v_c=175$~km$/$s in our fiducial model, but also
consider the effect of varying this threshold to $v_c=150$~km$/$s and
$v_c=200$~km$/$s in Figure~\ref{fig:result3}.  The left panel of
Figure~\ref{fig:ngamma} shows the $N_\gamma$ assignment for AGN in our
fiducial model ($v_c=175$~km$/$s) at redshifts $z=7$, 8, and 10.
Below the threshold mass, $N_\gamma\propto M_\mathrm{halo}$ and above
it $N_\gamma\propto M_\mathrm{bh}\propto M_\mathrm{halo}^{1.6}$,
following Equation~(\ref{eqn:bhmass}).  The middle panel of
Figure~\ref{fig:ngamma} shows $N_\gamma$ as a function of the halo
circular velocity.  We see that $N_\gamma$ sharply increases at
$v_c=175$~km$/$s.  This velocity corresponds to a different halo mass
at each of the three redshifts considered here, as seen in the right
panel of Figure~\ref{fig:ngamma}.  The magnitude of $N_\gamma$
increases with redshift to compensate for the decreasing number
density of haloes above the velocity threshold.  As described in
Section~\ref{sec:model}, where we discuss our calibration procedure,
the required total ionizing emissivity is dictated by our chosen
reionization model.

\bsp
\label{lastpage}
\end{document}